\documentclass[prb,aps,amsmath,amssymb,twocolumn]{revtex4-2}
\usepackage{graphicx,amsmath,amssymb,color,xcolor}
\usepackage[utf8]{inputenc}
\usepackage{ulem}
\usepackage{nicefrac}
\usepackage{multirow, makecell, ragged2e}
\usepackage[titletoc,title]{appendix}
\usepackage{physics}
\usepackage[colorlinks,bookmarks=true,citecolor=blue,linkcolor=red,urlcolor=blue]{hyperref}
\usepackage{cancel}

\newcommand{\be}{\begin{equation}}
\newcommand{\ee}{\end{equation}}

\newcommand{\ba}{\begin{eqnarray}}
\newcommand{\ea}{\end{eqnarray}}

\newcommand{\LLL}{\text{P}_{\text{LLL}}}

\renewcommand{\vec}[1]{\mbox{\boldmath$#1$}}

\def\beq{\begin{eqnarray}}
\def\eeq{\end{eqnarray}}

\newcommand{\nunp}{\frac{n}{2np+1}}

\begin{document}

\title{STM in the fractional quantum Hall effect: Spectroscopy of composite-fermion bound states}
\author{Mytraya Gattu$^1$, G. J. Sreejith$^{2}$, and J. K. Jain$^1$}
\affiliation{$^1$Department of Physics, 104 Davey Lab, Pennsylvania State University, University Park, Pennsylvania 16802,USA}
\affiliation{$^2$Indian Institute of Science Education and Research, Pune, India 411008}
\date{\today}

\begin{abstract}
The fractional quantum Hall states are non-Fermi liquids of electrons, in that their ground states and low energy excitations are described not in terms of electrons but in terms of composite fermions which are bound states of electrons and $2p$ quantized vortices. An electron or a hole at filling factor $\nu=n/(2pn+1)$, where $p,n$ are integers, is a complex molecule of $2pn+ 1$ quasiparticles (excited composite fermions) or quasiholes (missing composite fermions) and has its own internal excitations. Recent scanning tunneling microscopy experiments have succeeded in measuring the electron spectral functions of these states, which provides valuable information on the nature of these strongly correlated molecules and thereby on the short-distance correlations in the fractional quantum Hall liquids. These experiments exhibit several sharp peaks in the tunneling spectra. Detailed calculations based on the composite-fermion theory demonstrate multiple peaks in the local density of states, and we argue that the separation between the peaks represents interaction-corrected composite-fermion cyclotron energy. We discuss what aspects of experiments are explained by our model and which ones remain to be explained.
\end{abstract}
\maketitle

Recent breakthrough in performing scanning tunneling microscopy (STM) on fractional quantum Hall (FQH) liquids in graphene~\cite{Hu23,Farahi23} promises valuable new insights into the microscopic structure of the FQH states~\cite{Tsui82}. These measurements are made possible by the direct access to graphene, in contrast to the quantum wells which are embedded deep in GaAs heterostructures. The STM measurements can provide fundamental insights into the FQH phases.

The FQH states are ``non-Fermi liquids'' of electrons, in that they are described not in terms of electrons but rather in terms of composite fermions (CFs)~\cite{Jain89,Jain89b,Jain07,Halperin20}, which are bound states of electrons and an even number ($2p$) of quantized vortices, often pictured as bound states of electrons and $2p$ flux quanta, where a flux quantum is defined as $\Phi_0=hc/e$.  The CFs are distinct from electrons, as most readily evident from the fact that they experience a reduced magnetic field $B^*=B-2p\rho\Phi_0$ where $\rho$ is the electron (or CF) density.  With reference to an incompressible ground state which has an integer number of filled CF Landau levels (called $\Lambda$ levels), a CF excited to a higher $\Lambda$ level (the hole it leaves behind) is referred to as a quasiparticle (quasihole). As expected from general considerations~\cite{Laughlin83}, these quasiparticles / quasiholes carry a fractional charge~\cite{Jeon03b,Jeon04}.  The addition (removal) of an electron into (from) a FQH state at filling factor $\nu=n/(2pn+1)$ yields a complex bound state which is a superposition of states with $2pn +1$ quasiparticles (quasiholes) dressed by additional  CF-particle hole excitations out of the ground state. The STM experiments are a spectroscopic probe of the internal excitations of this bound state. A zeroth-order theoretical study of the spectral function predicted that in spite of the non-Fermi liquid nature of the FQH state, an electron or a hole exists as a sharp excitation~\cite{Peterson05}. The recent STM experiments~\cite{Hu23} observe more intricate additional structures, which has motivated the present study.

The present study crucially builds on the fact that the CF theory provides an excellent quantitative account of the excitations of the FQH liquids~\cite{Jain89,Jain90,Jain89b}. Section I of the Supplementary Material (SM)~\cite{SM-Gattu-23} shows the accuracy of the description of quasiparticles / quasiholes at $\nu=1/3$, 2/5 and 3/7. Extensive studies have demonstrated (see, for example, Ref.~\cite{Balram13}) that the CF theory provides a correlated basis of states that produces accurate approximations for the eigenstates of interacting electrons in the FQH regime, and the systematic inclusion of successively higher CF kinetic energy (CFKE) states into this basis yields successively higher interaction energy states of electrons. This is accomplished by diagonalizing the Coulomb interaction in the CF basis (which in general contains states that are not orthogonal or linearly independent), referred to as the method of CF diagonalization (CFD)~\cite{Mandal02}. We obtain the electron spectral function by performing CFD within a sufficiently large CF basis that allows us to reliably identify the resonant energy levels as well as their tunneling spectral weights. In the present study we assume that the physics arises entirely from a single Landau level (LL) which is equivalent to the spin-polarized lowest Landau level (LLL) of GaAs; we thus neglect any form factors arising from hybridization of spin, valley or layer degrees of freedom~\cite{McCann06}. We do not include the effect of any tip induced deformation of the FQH liquid and assume that the electron tunnels into a point in a translation invariant region of the FQH state. Our study also does not consider inelastic tunneling processes involving additional degrees of freedom, such as phonons, external to the FQH system.

In the constant height mode, STM measures the tunneling spectral function $A(E)$ given by the sum of two terms $A^+(E)$ and $A^-(E)$ representing electron and hole tunneling into the sample:
\begin{align}
&A^+(E)=\sum_m \left|\langle m,N+1 |c^\dagger|0,N \rangle\right|^2\delta(E-E_m^{N+1}+E_0^{N})\nonumber\\
&A^-(E)=\sum_m \left|\langle m,N-1 |c|0,N \rangle\right|^2\delta(E-E_0^{N}+E_m^{N-1})
\end{align}
where $c^{\dagger}$ creates an electron in the LLL at a point,
 $|0,N \rangle$ is the incompressible ground state of $N$ particles and $|m,N \rangle$ represents the $m^{\rm th}$ eigenstate of the system with $N$ particles. $E_m^N$ are the energies of the eigenstates of $N$ electrons labeled by $m$. For an electron (hole) added to an incompressible state of $N$ particles, we use  $E_{0}$ to refer to the ground state energy $E_{0}^{N}$.  The STM measures the energies relative to the chemical potential $\mu$, which is given, in the thermodynamic limit, by the energy per particle of the incompressible ground state (including the interaction with the background). We present our results as a function of $E-\mu$ below.

We will evaluate the spectral function in the spherical geometry \cite{Haldane83}, which minimizes finite size effect by eliminating the edge. We consider $N$ electrons on the surface of a sphere with a radial magnetic field arising from a fictitious magnetic monopole at the origin that produces a uniform radial field $B$ with a total magnetic flux $2Q\Phi_0$. The kinetic energy of the electrons is quantized into LLs with orbitals in the $n^{\rm th}$ LL ($n=0,1,\dots$) forming an angular momentum multiplet of angular momentum quantum number $l=Q+n$ and can be labeled by the $L_z$ quantum number $-l\leq m\leq l$. The single particle orbitals in the $n^{\rm th}$ LL are given by the monopole harmonics $Y_{Q,l,m}(\Omega)$\cite{Wu76,Wu77} where $\Omega=(\theta,\phi)$ are the coordinates on the surface of the sphere. In particular, the $n=0$ LL orbitals are given by $Y_{Q,Q,m}\sim u^{Q+m}v^{Q-m}$ where $u = \cos (\theta/2)\exp(\imath \phi/2)$ and $v = \sin (\theta/2)\exp(-\imath \phi/2)$. Jain's CF wave function for the incompressible state at filling $\nu=n/(2pn+1)$ is given by~\cite{Jain89,Jain90,Jain89b}
\begin{equation}
\label{eq:cf-sphere-definition}
\Psi_{\nu=\frac{n}{2pn+1}}=\LLL \phi_{n} \prod_{i<j}\left(u_{i}v_{j}-v_{i}u_{j}\right)^{2p}
\end{equation}
where $\phi_{n}$ is the Slater determinant corresponding to incompressible integer quantum Hall (IQH) state with $n$ filled LLs on a sphere with $2Q^{\star}=2Q-2p(N-1)$ flux passing through it. 
$\LLL$ projects the wavefunction into the $n=0$ LL, which can be implemented by the method in Refs.~\cite{Jain97,Jain97b}. A single quasihole (quasiparticle) can be constructed by replacing $\phi_n$ by a state containing a single hole in the $n$th LL (particle in the $n+1$th LL). Neutral excitons are made of such quasiparticle-quasihole pairs. These wave functions are known to represent the actual Coulomb states with a high degree of accuracy.

\begin{figure}
\includegraphics[width=0.22\textwidth]{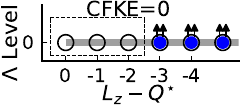}
    \includegraphics[width=0.22\textwidth]{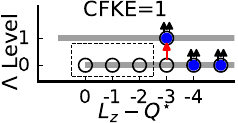}
\caption{Left panel: The minimal basis for a hole at $\nu=1/3$ enclosed by a dashed box. Right panel: A typical configuration of a basis state with unit CF kinetic energy (CFKE). Here and in the following figures the horizontal lines represent the $\Lambda$ levels, the available CF orbitals are shown by open circles, and the occupied orbitals have CFs, shown as blue dots decorated by two flux quanta (arrows). The spherical geometry is assumed, where $L_z$ is the angular momentum of the orbital and $Q^*$ is the monopole strength experienced by CFs. 
\label{fig:minimal-model-hole-1-3}
}
\end{figure}

\begin{figure}
    \includegraphics[width=0.22\textwidth]{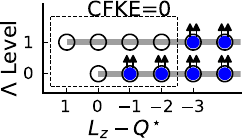}
    \includegraphics[width=0.22\textwidth]{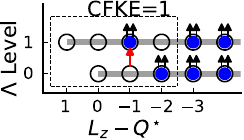}
    \includegraphics[width=0.22\textwidth]{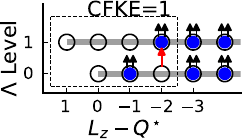}
    \includegraphics[width=0.22\textwidth]{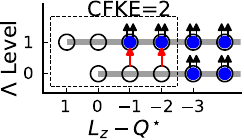}
    \includegraphics[width=0.255\textwidth]{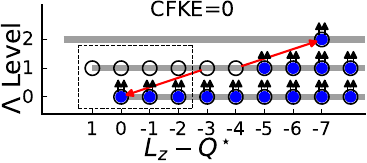}
    \includegraphics[width=0.22\textwidth]{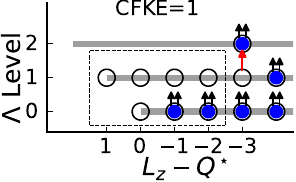}
        \caption{The top four panels depict the minimal basis that fully captures the hole added to the incompressible state at $\nu = \frac{2}{5}$. The red arrows indicate how these states are obtained starting from the minimum energy state (top left). All excitations are confined to a localized region depicted by the dashed box. The bottom two panels give examples of CF basis states beyond the minimal model. Their inclusion in CFD broadens the delta function peaks of the minimal model.  
        }
        \label{fig:minimal-model-hole-2-5}
\end{figure}

{\it Hole.} Tunneling of an electron out of the FQH system creates a hole.
For the Jain fractions $\nu=n/(2pn+1)$, the state with a hole at a point, say the north pole $\Omega=\omega=(u=1,v=0)$ with quantum numbers $L=Q,L_z=-Q$, is given by 
\begin{eqnarray}
&&c_{\omega} \Psi_{n\over 2pn+1}(\Omega_1, \cdots, \Omega_N)  \propto  \Psi_{n\over 2pn+1}(\Omega_1, \cdots, \Omega_{N-1},\omega) \nonumber \\
&=& \LLL \prod_{i<j=1}^{N-1}\left(u_{i}v_{j}-v_{i}u_{j}\right)^{2p}  \phi_n(\Omega_1, \cdots, \Omega_{N-1},\omega) \prod_{j=1}^{N-1} v^{2p}_j\nonumber
\end{eqnarray}
where $c$ represents the electron annihilation operator. 
This the CF form, i.e., it is a wave function multiplied by a Jastrow factor. It can be represented exactly as the linear combination of a finite number of simple CF states, referred to as the minimal basis, with $2pn+1$ quasiholes clustered near the origin (see Sec VI of SM~\cite{SM-Gattu-23}). The minimal basis for $\nu=1/3$ and $\nu=2/5$ is schematically shown in Figs. \ref{fig:minimal-model-hole-1-3} and  \ref{fig:minimal-model-hole-2-5} (the explicit basis function can be constructed in the standard manner -- by writing the corresponding IQHE wave function, multiplying by the Jastrow fact, and then projecting into the LLL), along with the CF kinetic energy (CFKE) for each basis function measured relative to the minimum CFKE basis function.  (At $\nu=2/5$, one combination of these states occurs at a different $L$ than the hole, leaving only three basis functions.) These figures also show how a larger CF basis can be constructed by adding excitons. The dimension of the minimal basis is independent of $N$ but increases rapidly with $n$ along the Jain sequence $\nu=n/(2pn +1)$ (Sec. VI of SM~\cite{SM-Gattu-23}).
  
\begin{figure}
 \includegraphics[height=0.24\textheight,width=0.48\textwidth]
    {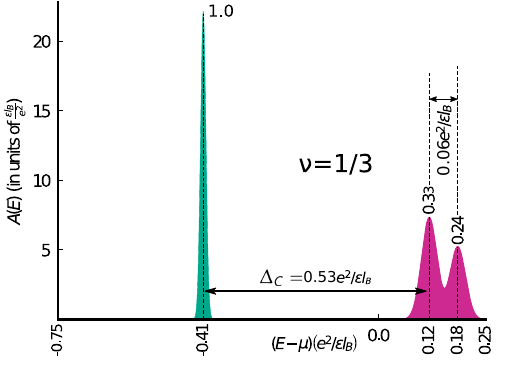}
    \includegraphics[height=0.24\textheight,width=0.48\textwidth]{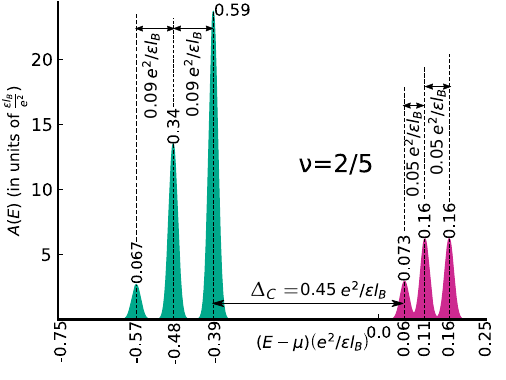}
    \includegraphics[height=0.24\textheight,width=0.48\textwidth]{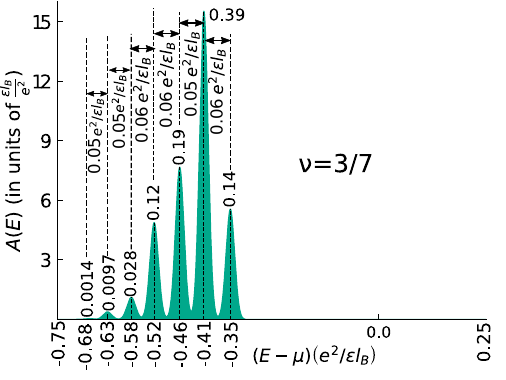}
    \caption{The spectral function $A(E)$ for FQHE at filling fraction $\nu = \frac{1}{3}, \frac{2}{5}$ and $\frac{3}{7}$ calculated using CF theory.
The chemical potential is denoted by $\mu$. The spectral function for $E>\mu$ ($E<\mu$) is shown  in purple (green). The spectral weight  under each peak is shown near its top; the peak positions and the spectral weights represent the thermodynamic values. All energies are quoted in units of $e^2/\epsilon l_B$. The heights of the peaks are proportional to the spectral weight under them.
    }
    \label{fig:spectral-function}
\end{figure}

We diagonalize the Coulomb interaction in the minimal basis using the method of CFD~\cite{Mandal02}, which produces approximate energy eigenstates with nonzero spectral weights. Comparison with exact diagonalization in small systems validates this minimal basis for spectral function calculation (Sec. IV of SM~\cite{SM-Gattu-23}). We also perform CFD within an enlarged basis also containing states with additional excitons and find that these also do not produce new peaks but, for cases where additional exciton basis functions are available with the same CFKE as the minimal basis function, causes a broadening of the peaks. The hole peak at $\nu=1/3$ is not broadened.

\begin{figure}
    \includegraphics[width=0.22\textwidth]{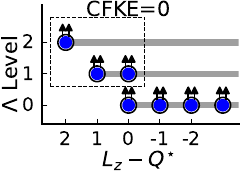}
    \includegraphics[width=0.22\textwidth]{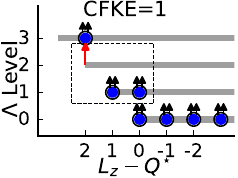}
    \includegraphics[width=0.22\textwidth]{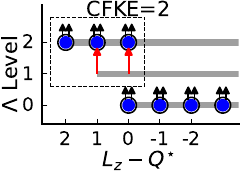}
    \caption{Examples of CF configurations used to represent an electron at $\nu = \frac{1}{3}$. The dotted rectangle shows the area in which the added electron is localized. 
    }  
    \label{fig:particle-1-3-basis}
\end{figure}

Fig.~\ref{fig:spectral-function} shows the spectral function on the hole side ($E<\mu$) shaded in green. The energy of each peak and its integrated spectral weight are shown on the figure; these numbers reflect the thermodynamic limits obtained from the CF theory (Sec. IV of SM~\cite{SM-Gattu-23}). The spectral weights add to unity. The line shapes of the peaks are schematic; for some cases, the line shapes obtained from CFD are shown in Sec. IV of SM~\cite{SM-Gattu-23}.

{\it Electron.} The state with an electron added at the north pole (with quantum numbers $L=L_z=Q$) is given by \cite{Peterson05}:
\begin{eqnarray}
\label{eq:electronAddedState}
&&c^{\dagger} (\omega) \Psi_{n\over 2pn+1}(\Omega_1,... \Omega_N) \nonumber \\
& &\propto {\rm A}[Y_{Q,Q,Q}(\Omega_{N+1}=\omega) \Psi_{n\over 2pn+1}(\Omega_{1}\dots \Omega_{N})].
\end{eqnarray}
This does not have a CF form, because it is the LLL projection of 
$A Y_{Q,Q,Q}(\Omega_{N+1}) \phi_{n}(\Omega_1, \cdots, \Omega_N) \prod_{i<j=1}^{N}\left(u_{i}v_{j}-v_{i}u_{j}\right)^{2p} $
which does not have the proper Jastrow factor for all $N+1$ particles. 
As a result, in contrast to the hole, the added electron cannot be exactly represented as a linear superposition of states with a simple CF structure. We proceed by constructing a CF basis in which the lowest $n$ $\Lambda$ levels are full and $2n+1$ quasiparticles occupy excited $\Lambda$-level orbitals to produce the desired total angular momentum quantum numbers $L=L_z=Q$. Let us illustrate with the example of $\nu=1/3$, shown in Fig.~\ref{fig:particle-1-3-basis}. Though there are an infinite number of such states, we can place a cut-off on the net CFKE and the highest $\Lambda$ level of these particles to obtain a finite number of such basis states.  A basis containing $16$ CF configurations with the correct quantum numbers is obtained if we have a net CFKE of $0,1$ and $2$ and the three quasiparticles are allowed to be in $\Lambda$ levels $1,2$ and $3$.  (We expect that excitons far from the north pole will have negligible overlap with the electron.) Extrapolating to the thermodynamic limit, the finite CF basis captures (see Fig.~\ref{fig:spectral-function}) $\sim$76\% of the spectral weight for $\nu=1/3$ and $\sim$ 35\% at $\nu=2/5$ (see Fig.~\eqref{fig:spectral-function}). At $\nu=3/7$ we expect a much smaller number whose precise determination is possible but computationally expensive. We have found that increasing the dimension of the CF basis by a factor of two does not significantly change the total weight, which suggests that the electron is not fully representable in terms of CFs. We speculate (on the basis of finite system studies) that the remainder of the spectral weight is distributed in a high energy tail. This indicates a fundamental asymmetry between the additions of a hole and an electron in a FQH system.

{\it Qualitative understanding of the results.} The spectral functions shown in Fig.~\ref{fig:spectral-function} represent the principal result of our study. To gain insight into its features, we consider a model where the CFs are taken as noninteracting. Within this model the energies can be expressed in terms of the $\nu$ dependent CF cyclotron energy $\hbar\omega^*_{\nu}$. For example, the separation between the first electron and the first hole peaks is given by  $4\hbar\omega^*_{1/3}$ at $\nu=1/3$ and $8\hbar\omega^*_{2/5}$ for $\nu=2/5$, and the separation between the peaks {within} the electron or the hole spectral function is $\hbar\omega^*_{\nu}$. While this captures the qualitative features, all these energies are renormalized by the inter-CF interaction, which is significant here given the physical proximity of the excited CFs. The peaks are approximately equally spaced for the electron and also for the hole, making it meaningful to identify the separation with a renormalized $\hbar\omega^*_{\nu}$. However, the renormalized $\hbar\omega^*_{\nu}$'s for the electron and the hole sides are not equal at a given $\nu$. On intuitive grounds, one expects that the CF cyclotron energy on the hole side should be larger than that on the electron side because the local $B^*$ for the hole is larger due to the reduced density; this is consistent with the behavior seen in Fig.~\ref{fig:spectral-function}. The multiple peaks thus reflect the $\Lambda$ level structure that is renormalized by the residual interaction between the CFs. This understanding can be extended to other fractions not accessible to detailed study.

{\it $\nu=n/(2n-1)$ FQH states.} On account of the particle hole symmetry of the problem within the a LL, the spectral function for the electron (hole) tunneling into a state at $\nu=1-n/(2pn+1)$ is identical to that for the hole (electron) tunneling into a state at filling $\nu=n/(2pn+1)$, modulo a rigid shift in the energy.

{\it Comparison with experiment.} A number of aspects that might be relevant in experiments are not included in our model. The effects of disorder and LL mixing, screening of the interaction by the backgate, and the role of spin are neglected, and it is assumed that the influence of the STM tip's potential on the FQH state is insignificant. With this caveat, let us consider how the above results compare to the experiment of Ref.~\cite{Hu23}. 

In Ref.~\cite{Hu23} the behavior for hole (electron) at $n/(2n+1)$ does not match with that for electron (hole) at $1-n/(2n+1)$. Much more structure is seen for $\nu>1/2$ than for $\nu<1/2$. This underscores the importance of LL mixing, which breaks particle-hole symmetry. Why LL mixing is more significant for $\nu<1/2$ is an important question, which very likely involves subtle physics that is beyond the scope of the current study.

Experiments do see sharp peaks, as expected from above discussion and from previous studies~\cite{Peterson05}. On the electron side of $\nu=2/3$, there is a sharp peak (see Fig.~3B. of Ref.~\cite{Hu23}), which we identify with the hole partner of the single peak on the hole side at $\nu=1/3$ (Fig.~\ref{fig:spectral-function}). Additional structure is seen on the electron side of $\nu=2/3$ including a broad peak, the origin of which is unclear. On the hole side of $2/3$ there is a sharp peak with a shoulder, which could be two unresolved peaks, as expected from Fig.~\ref{fig:spectral-function}. For the hole side of $2/5$, or the electron side of 3/5, we expect three peaks, which may be consistent with experiments (Fig.~3B. of Ref.~\cite{Hu23}). On the hole side of 3/5, our study predicts three peaks with a small weight; experimentally a broad peak is seen with a smaller weight. 

For a more quantitative comparison, we note that the separation between the closest electron and hole peaks is approximately 0.53 and 0.45 $e^2/\epsilon l_B$ for $\nu=1/3$ and $\nu=2/5$. Assuming $\epsilon=3.5$ and $B=14$T, this translates into 29 meV and 24 meV. Experimentally, the separation is $\sim$ 16 meV for $\nu=2/3$ and $\sim$ 12 meV for $\nu=3/5$. The separation between the peaks on the electron side of $\nu=3/5$ is on the order of $\sim$ 3 meV in experiments, which corresponds to 0.054 $e^2/\epsilon l_B$. This is to be compared to the theoretical separation of 0.09 $e^2/\epsilon l_B$.  While the origin of the factor of $\sim$2 discrepancy between the theoretical and experimental gaps is not known at present, we note that a similar level of deviation has been seen for various gaps for the FQHE states in GaAs, often attributed to LL mixing and disorder. At this stage, it is not possible to tell in experiments how the renormalized $\hbar\omega^*_{\nu}$'s differ on the hole and the electron sides.

Theoretically, we expect weaker peaks on the electron side of $\nu=n/(2n+1)$ or the hole side of $\nu=1-n/(2n+1)$. This appears to be the case as seen in Fig.~3A of Ref.~\cite{Hu23}.  

In summary, we have performed a detailed theoretical study of the spectral function in the FQHE and found good qualitative and reasonable quantitative agreement with many features of the experiments of Ref.~\cite{Hu23}. We have identified aspects that are not well understood. We note that while the presence of multiple peaks arises fundamentally from the fractionalization of an electron into an odd number of CFs (in that they reflect the internal excitations of their electron-like bound state), it cannot be used to deduce the fractional braiding statistics of the quasiparticles / quasiholes~\cite{Halperin84,Arovas84}. The braiding statistics is well defined only when the separation between quasiparticles / quasiholes is large compared to their size~\cite{Kjonsberg99,Kjonsberg99b,Jeon03b,Jeon04}, in contrast to the situation here which is dominated by configurations wherein the quasiparticles / quasiholes are all crowded in a small region. It has been proposed that disorder mediated tunneling can help reveal the fractional statistics through STM~\cite{Papic18} and that the STM signals contain signatures of entanglement in the FQH phase~\cite{Pu22}.

{\it Note added.} While finishing our manuscript, we became aware of an independent study by Pu {\it et al.}~\cite{Pu23} which has a significant overlap with our work.

We are grateful to Ali Yazdani for numerous insightful discussions on STM of FQHE, which motivated the present work. We thank Ajit C. Balram, Songyang Pu, and Z. Papic for pointing out certain quantitative errors in the spectral weights in Ref.~\cite{Peterson05}, which we have also confirmed.  M.G. and J.K.J. acknowledge financial support from the U.S. National Science Foundation under grant no. DMR-2037990. G. J. S thanks Condensed Matter Theory Center and Joint Quantum Institute, University of Maryland, College Park for their hospitality during the completion of this work, and  Ashish Arora and Biswajit Karmakar for useful discussions.

\bibliography{biblio_fqhe.bib}

\pagebreak

{\bf Supplementary material for\\ STM in the fractional quantum Hall effect: Spectroscopy of composite-fermion bound states}

\section{Quasiparticles and quasiholes\label{supp_sec:QPQH}}
In the calculations presented in this work, we describe the state after a hole (electron) has been added to the fractional quantum Hall (FQH) state as a collection of closely packed quasiholes (quasiparticles). Therefore, a quantitatively accurate description of the quasihole (quasiparticle) states is crucial to precisely determine the spectral function.

By comparing with exact diagonalization (ED) results for the Coulomb interaction, we explictly demonstrate that CF theory provides a highly accurate quantitative description for a single quasiparticle and quasihole at $\nu = \frac{1}{3}, \frac{2}{5}$ and $\nu = \frac{3}{7}$.

We use the spherical geometry as we are interested only in bulk excitations~\cite{Haldane83}. We consider $N$ electrons on the surface of a sphere with a radial magnetic field arising from a fictitious magnetic monopole. The monopole produces a uniform radial magnetic field $B$ with a total magnetic flux $2Q\Phi_0$ where $\Phi_0$ is the magnetic flux quantum. The kinetic energy of the electrons is quantized into Landau levels(LLs) with orbitals in the $n^{\rm th}$ LL ($n=0,1,\dots$) forming an angular momentum multiplet of angular momentum quantum number $l=Q+n$. The different states in each multiplet can be labeled by the $L_z$ quantum number $-l\leq m\leq l$. The single particle wave functions in the $n^{\rm th}$ LL are given by the monopole harmonics $Y_{Q,l,m}(\theta,\phi)$~\cite{Wu76,Wu77} where $\theta,\phi$ are the coordinates on the sphere. In particular, the $n=0$ LL orbitals are given by $Y_{Q,Q,m}\sim u^{Q+m}v^{Q-m}$ where $u = \cos (\theta/2)\exp(\imath \phi/2)$ and $v = \sin (\theta/2)\exp(-\imath \phi/2)$. Jain's CF wave function for the incompressible state at filling $\nu=n/(2pn+1)$ is given by~\cite{Jain89,Jain90,Jain89b}
\begin{equation}
\label{eq:cf-sphere-definition}
\Psi_{\nu=\frac{n}{2pn+1}}=\LLL \phi_{n} \prod_{i<j}\left(u_{i}v_{j}-v_{i}u_{j}\right)^{2p}
\end{equation}
where $\phi_{n}$ is the Slater determinant corresponding to the incompressible integer quantum Hall (IQH) state with $n$ filled LLs on a sphere with $2Q^{\star}=2Q-2p(N-1)$ flux passing through it. 
$\LLL$ projects the wave function into the lowest LL (LLL). 
\begin{figure}
\includegraphics[width=.9\columnwidth]{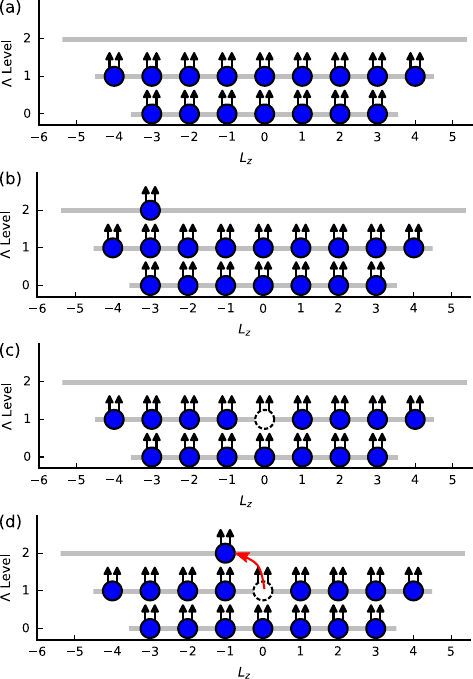}
\caption{Schematic representation of the occupancy of the CF orbitals in the incompressible state (a), the quasiparticle state (b), the quasihole state (c), and the neutral exciton (d) of the $2/5$ FQH phase.\label{supp_fig:cfBasics}
}
\end{figure}

\begin{figure}
\includegraphics[width=\columnwidth]{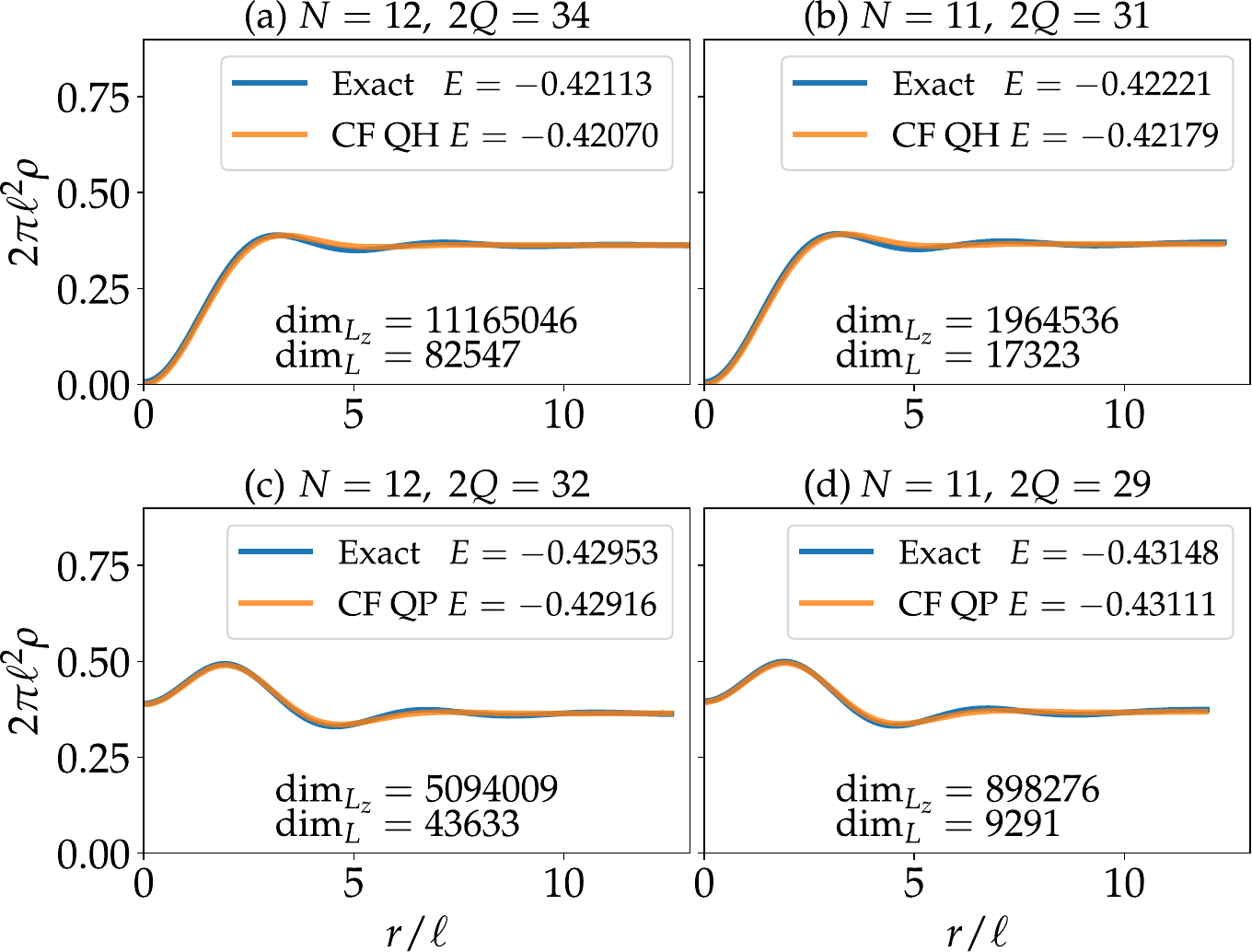}
\caption{(a,b) Electron density $\rho$ of the CF quasihole at $\nu=\frac{1}{3}$ (orange) compared with the density of the exact quasihole (blue). The Coulomb energy $E$ per particle (in units of $\frac{e^{2}}{\varepsilon \ell}$) for the exact and CF quasiholes are shown in the figures. Panels (a) and (b) show two different system sizes. The $x$-axis represents the arc distance from the north pole along a longitude. The quantity $l$ is the magnetic length. The symbol ${\rm dim}_{L_{z}}$ denotes the dimension of the Fock space of electrons at $L_{z}=-Q^{\star}$. The symbol ${\rm dim}_{L}$ denotes the dimension of the Fock space at $L=Q^{\star}$.
(c,d) Electron density $\rho$ of the CF quasiparticle at $\nu=\frac{1}{3}$ (orange) compared with the density of the exact quasiparticle (blue). The Coulomb energy $E$ per particle (in units of $\frac{e^{2}}{\varepsilon \ell}$) for the exact and CF quasiparticles are shown in the figures. Panels (a) and (b) show two different system sizes. The symbol ${\rm dim}_{L_{z}}$ denotes the dimension of the Fock space of electrons at $L_{z}=Q^{\star}$.
\label{supp_fig:oneThird}
}
\end{figure}
A schematic representation of the occupancy of the CF orbitals in the incompressible states $\psi_{\nu = \frac{n}{2np+1}}$ is shown in Fig.~\ref{supp_fig:cfBasics}(a).
We can construct a single quasihole (quasiparticle) of the FQH state by replacing $\phi_{n}$ with its quasihole (quasiparticle), i.e., the state containing a quasihole(quasiparticle) of the $\nu=n$ integer quantum Hall effect. These are shown schematically in Fig.~\ref{supp_fig:cfBasics}(b,c).
Neutral excitons are made of such quasihole-quasiparticle pairs (Schematically shown in Fig.~\ref{supp_fig:cfBasics}(d)).

\begin{figure}
\includegraphics[width=\columnwidth]{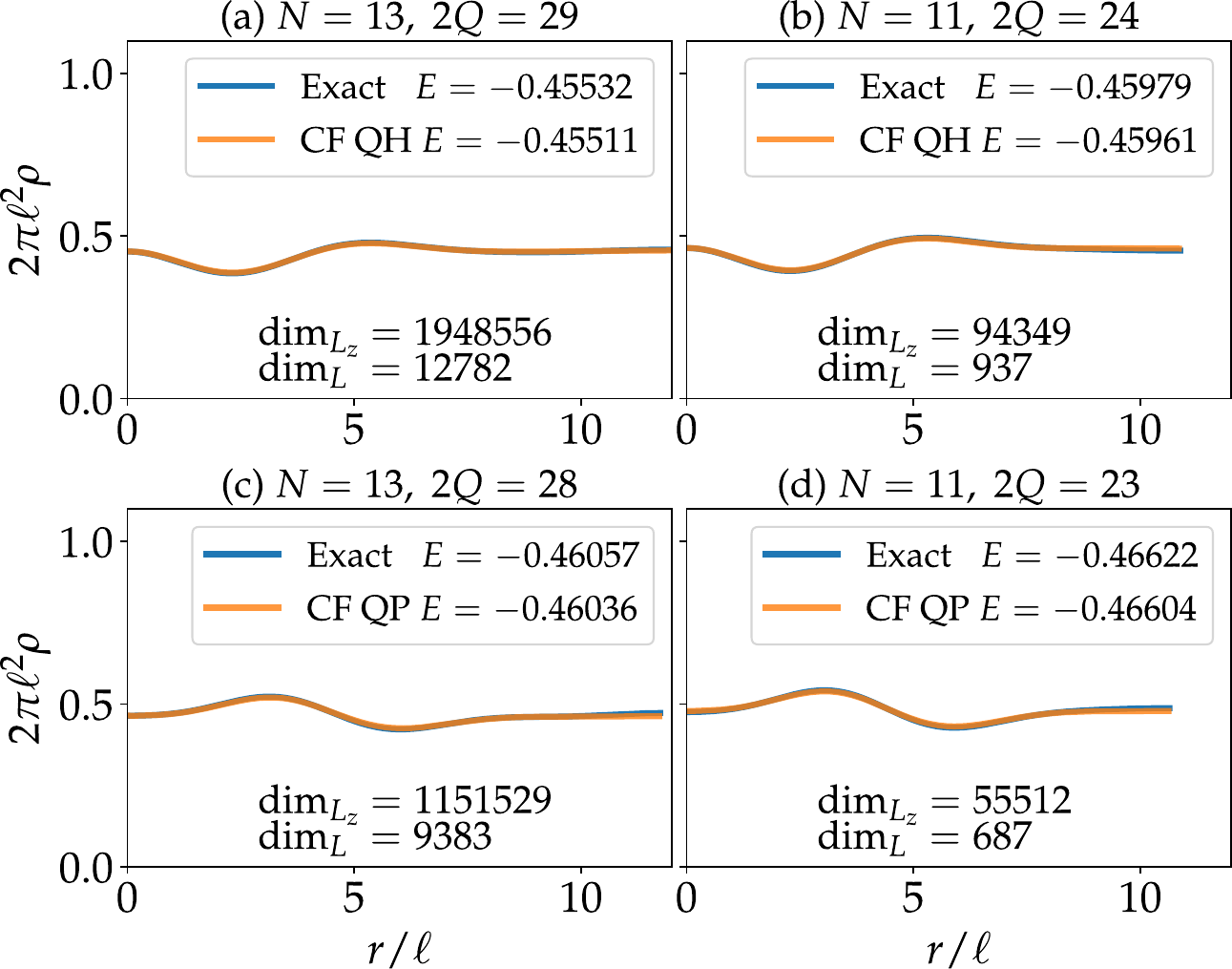}
\caption{(a,b) Electron density $\rho$ of the CF quasihole at $\nu=\frac{2}{5}$ (orange) compared with the density of the exact quasihole (blue). 
The symbol ${\rm dim}_{L_{z}}$ denotes the dimension of the Fock space of electrons at $L_{z}=-Q^{\star}-1$. The symbol ${\rm dim}_{L}$ denotes the dimension of the Fock space at $L=Q^{\star}+1$.
(c,d) Electron density $\rho$ of the CF quasiparticle at $\nu=\frac{2}{5}$ (orange) compared with the density of the exact quasiparticle (blue). 
 The symbol ${\rm dim}_{L_{z}}$ denotes the dimension of the Fock space of electrons at $L_{z}=Q^{\star}+1$.
\label{supp_fig:2Fifths}
}
\end{figure}

\begin{figure}
\includegraphics[width=\columnwidth]{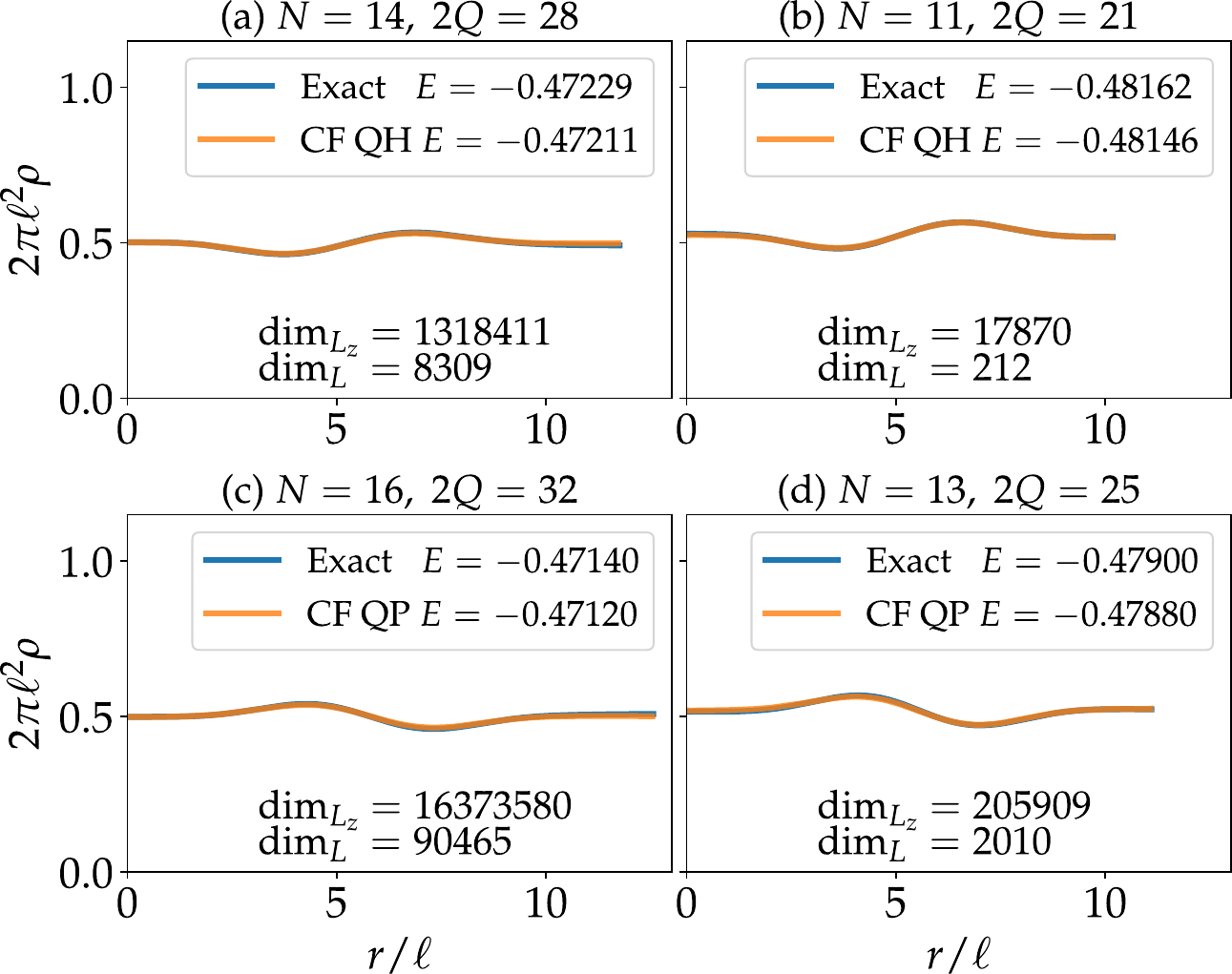}
\caption{(a,b) Electron density $\rho$ of the CF quasihole at $\nu=\frac{3}{7}$ (orange) compared with the density of the exact quasihole (blue). 
The symbol ${\rm dim}_{L_{z}}$ denotes the dimension of the Fock space of electrons at $L_{z}=-Q^{\star}-2$. The symbol ${\rm dim}_{L}$ denotes the dimension of the Fock space at $L=Q^{\star}+2$.
(c,d) Electron density $\rho$ of the CF quasiparticle at $\nu=\frac{3}{7}$ (orange) compared with the density of the exact quasiparticle (blue). 
 The symbol ${\rm dim}_{L_{z}}$ denotes the dimension of the Fock space of electrons at $L_{z}=Q^{\star}+2$. \label{supp_fig:3Sevenths}
}
\end{figure}

In the following, we consider a quasiparticle (QP) or a quasihole (QH) located at the north pole of the sphere.
In the CF wavefunction describing a quasihole of $\nu = n/(2pn+1)$ at the north pole, the highest weight orbital ($L=L_z=Q^*+n$) in the $(n-1)^{\rm th}$ $\Lambda$ level is empty in the otherwise fully occupied CF state of $n$ filled $\Lambda$ levels in the Slater determinant $\phi_{n}$.
In the CF wavefunction describing a quasiparticle of $\nu = n/(2pn+1)$ at the north pole, the highest weight orbital ($L=L_z=Q^*+n+1$) in the $n^{\rm th}$ $\Lambda$ level is occupied in addition to the first $n$ fully occupied $\Lambda$ levels in the Slater determinant $\phi_{n}$.
Figures \ref{supp_fig:oneThird}(a,b) show the electron density of the quasihole of the $\nu=1/3$ FQH state at the north pole calculated using CF theory. We compare this with the density of the lowest energy eigenstate in the exact spectrum of the Coulomb interaction at the same angular momentum quantum numbers at the two largest accessible system sizes.
The energy and the electron density of the quasihole calculated using CF theory and exact diagonalization (ED) of the Coulomb interaction agree closely. A similar comparison for the case of the quasiparticle at $\nu = 1/3$ is shown in Fig. \ref{supp_fig:oneThird}(c,d). Figures \ref{supp_fig:2Fifths} and \ref{supp_fig:3Sevenths} show the comparisons for the quasihole and quasiparticle of the $\nu=2/5$ and $\nu=3/7$ FQH states respectively.  For smaller $N$, such comparisons have been reported previously~\cite{Jain07}. Previous work has also demonstrated the accuracy of states containing several quasiparticles or quasiholes; see, for example~\cite{Balram13}.

\section{Addition of an electron or a hole\label{supp_sec:addingRemovingElectron}}
Having described the CF wave functions for the incompressible FQH states and their simplest excitations, we now discuss the construction of the wave function for the addition (removal) of an electron to (from) a point within the lowest LL.

A state $\Psi_{\nu+e}$ in which a maximally localized electron is added at a point $\Omega$ is given by $c^\dagger(\Omega)\Psi_{\nu}$ where $c^{\dagger}(\Omega) = \sqrt{4\pi}\sum_m\overline{Y}_{Q,Q,m}(\Omega)C^{\dagger}_{Q,Q,m}$ is the LLL projected electron creation operator. Here $C^{\dagger}_{Q,Q,m}$ adds an electron into the LLL angular momentum eigenstate $Y_{Q,Q,m}$. Translation invariance of the problem (which is manifested as the rotational invariance on the sphere) permits us to choose $\Omega$ to be the north pole where the electron added state has a particularly simple form~\cite{Peterson05}:
\begin{equation}\label{eq:app:electronAddedState}
\Psi_{\nu+e} = c^\dagger (\omega) \Psi_{\nu} = {\rm A}[Y_{Q,Q,Q}(\Omega_{N+1}) \Psi_{\nu}(\Omega_{1}\dots \Omega_{N})]
\end{equation}
where ${\rm A}$ represents the anti-symmetrization operation, $\Omega_{i}$ represents the $i^{\rm th}$ electron coordinate and the symbol $\omega\equiv (\theta=0,\phi)$ represents the north pole. 

The hole state with $N-1$ electrons is obtained by removal of the electron from a point $\omega$ on the sphere. Its wave function is given by 
\begin{equation}\label{eq:electronRemovedState}
\Psi_{\nu-e} = c(\omega)\Psi_\nu\propto \Psi_{\nunp}(\Omega_{1},\dots,\Omega_{N-1},\Omega_{N}=\omega)
\end{equation}
where one of the $N$ electron coordinates is replaced by $\omega$.

The incompressible state at $\nu = \frac{n}{2pn+1}$ (Eq.~\eqref{eq:cf-sphere-definition}) has a total angular momentum $L=0$. Thus, the addition or removal of the electron(from within the LLL) produces a state of angular momentum $L=Q$. If this electron is added/removed at the north pole, the resulting states $\Psi_{\nu\pm e}$ have a total azimuthal angular momentum quantum number of $L_z=\pm Q$. This follows from the fact that $c^\dagger(\omega)$ is the same as $C^{\dagger}_{Q,Q,m=Q}$ which adds an electron into the single particle state with $L=L_z=Q$.

\begin{figure*}
\includegraphics[width=0.4\textwidth]{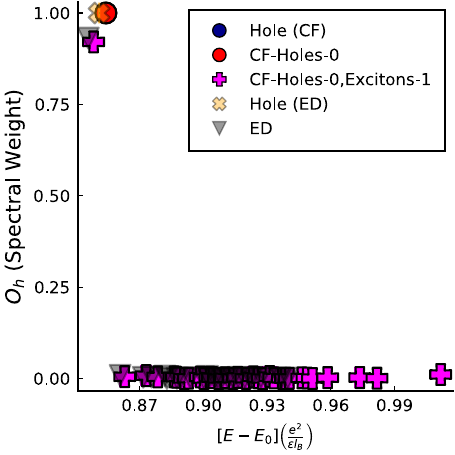}
\includegraphics[width=0.4\textwidth]{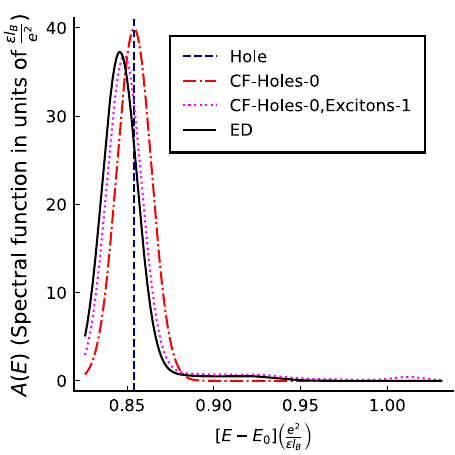}
\caption{Spectral weights of the hole added to an incompressible state at filling fraction $1/3$ calculated using different choices of basis in a system of $N=12, 2Q=36$. 
(left): $y$-axis shows the spectral weights $O_{h}$ and $x$-axis shows the energy of the eigenstates.
The red filled circle marked as `CF-Holes-$0$' in the legend show the spectral weight calculated using the minimal basis set which contains only one state which has a CFKE of 0 (Fig.~1 left panel of the main text).
Purple crosses labeled `CF-Holes-$0$+Excitons-$1$' show the spectral weights calculated by including additional CF states containing excitons, such as the state shown in Fig. 1 (right panel) of the main text. All such states have a CFKE of 1. 
Orange x-crosses shows the energy expectation value of the hole added to the Coulomb ground state obtained from ED. The spectral weights calculated from the ED eigenstates are shown as black triangles. All approaches show a single spectral peak which is not broadened by the excitons.
(right) Spectral function $A(E)$ on the $y$-axis obtained by a gaussian binning of the spectral weights shown in the left panel. The dotted vertical line shows the energy expectation value of the hole.
\label{fig:hole-1-3-ed-comparison}
}
\end{figure*}
\section{Calculation details}

We use the CF variational states described in Sec. \ref{supp_sec:QPQH} to estimate the spectral function. The incompressible ground state (Eq.~\eqref{eq:cf-sphere-definition}) from CF theory forms an excellent approximation for $|0, N \rangle$ and allows us to represent the electron and hole added states $c^\dagger|0, N \rangle$ and $c|0, N \rangle$ with the states described in Eqs.~\eqref{eq:app:electronAddedState}, \eqref{eq:electronRemovedState}.
Only those energy eigenstates $|m, N \rangle$, which have a finite overlap with electron/hole added states contribute to the spectral function.
An excellent approximation to these energy eigenstates can be obtained by a diagonalization of the Coulomb interaction (the Hamiltonian) within a basis of low energy CF excitations that are localized near the tunneling point and that have the correct quantum numbers. These states are shown in Figs.~1 and 2 of the main text.
The matrix elements required for the diagonalization of the Hamiltonian as well as the amplitudes $\langle m,N+1 |c^\dagger|0,N \rangle$ and  $\langle m,N-1 |c|0,N \rangle$ can be efficiently computed using Monte Carlo methods. In the diagonalization within the CF Hilbert space, we use the Metropolis-Hastings-Gibbs Monte Carlo scheme for the necessary overlap and Hamiltonian matrix element calculations and the Jain-Kamilla projection to approximate the $\LLL$ operator~\cite{Kamilla96} efficiently. The CF basis states are generally not orthogonal. We can obtain the eigenvalues and the eigenvectors of the Hamiltonian in a non-orthogonal basis by diagonalizing the matrix $O^{-1}H$, where $O$ is the overlap matrix and the Hamiltonian matrix $H$ in the space of the non-orthogonal basis states.
The CF basis states are by themselves not eigenstates of $L^2$. By rotating to the $L^2$ basis using the Clebsch-Gordan coefficients, we can block-diagonalize the overlap and Hamiltonian matrices before performing the diagonalization. In particular, this helps us identify the eigenstates in the $L=L_z=\pm Q$ block, which are relevant for the spectral function calculations.
The mean values and errors were estimated using the jack-knife method.
Spectral functions are calculated from the spectral weight by replacing the Dirac delta functions in Eq.~1 of the main text with normal distributions of width $0.01\frac{e^{2}}{\varepsilon l_{B}}$.
The spectral function can also be computed for small systems using eigenstates obtained from ED. We have retained the lowest $100-150$ eigenstates from ED in a basis of $L_{z} = \pm Q$ for the electron and hole, respectively. Out of these, we have only retained the states at the correct angular momentum $L=Q$ that can have a finite spectral weight with either the electron or the hole. We use these spectral weights to check the sufficiency of the finite variational bases employed for the hole and electron spectral functions containing only quasiholes and quasiparticles. 

Additionally, within the variational approach, we can perform a more accurate energy eigenstate calculation by using a larger variational space of the aforementioned finite-sized bases and additional excitons of low energy\footnote{Certain $L^2$ eigenstates containing excitons vanish upon $\LLL$ projection. These can be identified from the singular values of the overlap matrix. Such null vectors combinations are eliminated before calculating the energies}. The addition of the excitons results only in a broadening the peaks that we already found from the simpler finite-bases calculations, confirming their sufficiency. 

\begin{figure*}
    \includegraphics[width=0.4\textwidth]{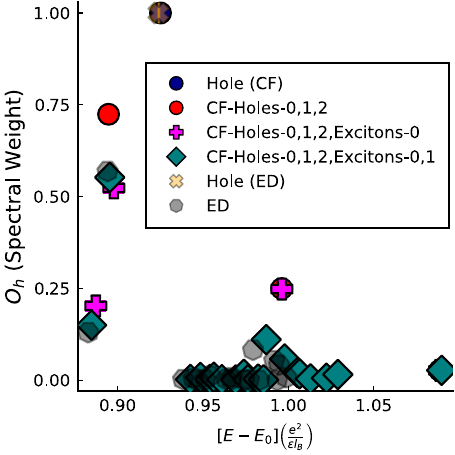}
    \includegraphics[width=0.4\textwidth]{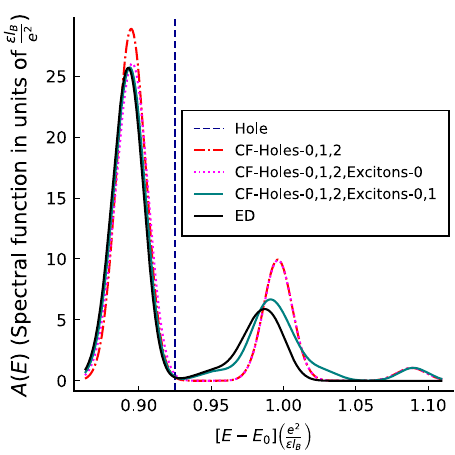}
    \caption{(left) The spectral weights of the hole added to $\nu=\frac{2}{5}$ as a function of the energies. 
    The system has $N=11$ particles at $L=Q=13$. 
    CF-Hole-$0,1,2$ basis refers to the minimal basis which contains states of CFKE $0$, $1$ and $2$ shown in Fig. 2 (top two rows) of the main text. 
    Inclusion of excitons with a total CFKE=$0$ and CFKE=$1$ in the variational space produces progressively better approximations to the ED spectral weights. 
    (right) The figure shows that the addition of excitons at different CFKEs only broadens the peaks without changing the qualitative features. \label{fig:hole-2-5-ed-comparison-1}}
\end{figure*}

\begin{figure*}
    \includegraphics[width=0.4\textwidth]{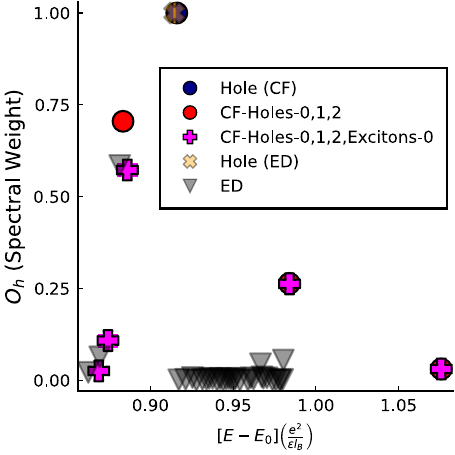}
    \includegraphics[width=0.4\textwidth]{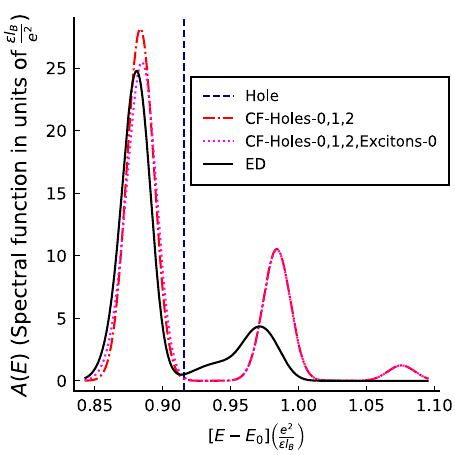}
    \caption{(left)The spectral weights of the hole added at $\nu=\frac{2}{5}$ as a function of the energies. 
    The system has $N=13$ particles at $L=Q=\frac{31}{2}$. 
    CF-Hole-$0,1,2$ basis refers to the minimal basis.
    We see that an inclusion of excitons of $0$ net CFKE is necessary to capture the Coulomb eigenstates (obtained using ED) underlying the lowest energy peak.
    (right) Again, a comparison of the spectral function shows that the addition of excitons at different CFKEs only broadens the peaks without changing the qualitative features.}
    \label{fig:hole-2-5-ed-comparison-2}
\end{figure*}

\section{Addition of a hole}\label{sec:hole-state}

This section provides a more complete account of the calculations leading to the thermodynamic results described in the main article.

The state obtained by the addition of a hole to a state at a Jain sequence filling fraction $\nu=n/(2pn+1)$ can be exactly represented as a linear combination of a finite number (that does not increase with system size) of CF states each containing $2pn+1$ quasiholes. In Sec. \ref{supp_sec:holeDecomposition}, we show that such a minimal basis set exists for all parallel flux-attached Jain fractions of incompressible states.
This basis provides an excellent approximation to the positions of the peaks and their spectral weights in the hole spectral function at each Jain sequence filling fraction $\nu=n/(2pn+1)$.

\subsection{$\nu=1/3$\label{sec:hole-13rd}}
The minimal basis for the case of $\nu=1/3$ contains only one state which exactly equals the state with a hole added. Energy expectation of this state therefore gives the location of the only peak in the hole spectral function which contains the entire spectral weight. This is shown in Fig. \ref{fig:hole-1-3-ed-comparison} (red filled circle in left panel) for the case of an $N=12$ particle system. The black filled triangles show how the spectral weights of the hole added to the exact Coulomb ground are distributed among the exact Coulomb eigenstates. The spectral weights from the ED calculation shows a single sharp spectral peak whose energy and strength are in agreement with the minimal basis.

To better approximate the energy eigenstates we consider a larger basis of states obtained by adding excitons to the minimal basis state. There are no such states with the same CFKE as the minimal basis state which is taken to have a CFKE of 0; the simplest of these exciton states have a CFKE of 1 (in units of CF cyclotron energy). We therefore expect the minimal basis state to not couple to the exciton degrees of freedom. Purple crosses (`$+$') symbols in the Fig.~\ref{fig:hole-1-3-ed-comparison} show the spectral weights calculated in this enlarged variational space. Consistent with the expectation, we find that excitons do not cause any broadening of the spectral peak. Figure \ref{fig:hole-1-3-ed-comparison} (right) shows the spectral function calculated using a gaussian binning of the spectral weights.
 \begin{figure}
    \includegraphics[width=0.4\textwidth]{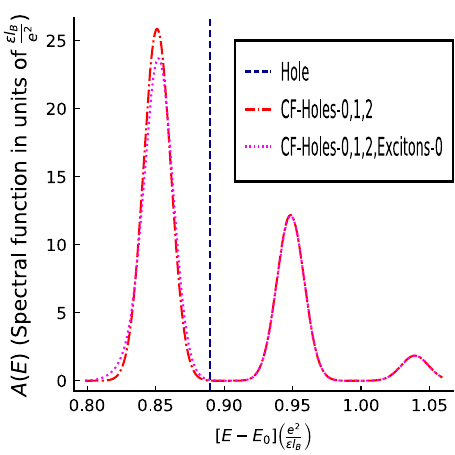}
    \caption{Spectrum function for the hole at $\nu=\frac{2}{5}$ in the largest system ($N=27,L=Q=33$) that we have studied showing that the inclusion of excitons (of CFKE $0$) only broadens the peaks from the minimal basis (CF-Holes-0,1,2).}
    \label{fig:hole-2-5-excitons-0-largest-size}
\end{figure}
\subsection{$\nu=2/5$}
The minimal basis that contains the entire spectral weight of the hole is shown in Fig. 2 (top two rows) of the main text. There are four states in this basis irrespective of the number of particles in the system. There are three (independent) states in this space with the angular momentum quantum numbers that match with the hole added state. Extending the observations from the case of $\nu=1/3$ we expect three peaks in the spectral function. 

Fig. \ref{fig:hole-2-5-ed-comparison-1} shows the hole spectral weights calculated using different bases. Of the three spectral weights calculated from the minimal basis (shown as filled red circular markers) the one at the largest energy has a very small spectral weight. The exact ED calculations show a larger number of states with finite but smaller spectral weights. However as shown in the spectral function obtained by binning the contributions of nearby energy eigenstates are the same for both the minimal basis and exact ED basis.

The finer details seen in ED can be reproduced by adding additional variational states to the minimal basis. First we added all single exciton states with the same $L_z=Q$ quantum number and with net CFKE of 0 (relative to the lowest CFKE minimal basis state). Figure 2 (bottom-left) of the main text shows one such state. Such states with CFKE=0 are possible here as the excess CFKE of the exciton can be compensated by the rearrangement of the the five holes near the north pole allowing an electron here to move from the second LL to the LLL. The spectral weights constructed from this basis set captures the finer details of the spectral weight and eigenstates in the lowest energy region. The two lowest energy states and their spectral weights are reproduced by the extended basis containing CFKE=0 exciton states.

We can now add the CFKE=1 states with single excitons to the basis set. While the CFKE=0 did not change the spectral weight of the second peak in the minimal basis, the CFKE=1 splits this spectral weight into several finer states thereby capturing the finer details in the ED spectral weights. 

As can be seen from the right panel of Fig. \ref{fig:hole-2-5-ed-comparison-1}, the binned spectral function continues to match the minimal basis even with the inclusion of further degrees of freedom. Figure  \ref{fig:hole-2-5-ed-comparison-2} shows similar results for a slightly larger system of $N=13$ particles.
We have also confirmed that this remains true in the thermodynamic limit by comparing the spectral functions obtained from the minimal basis and the minimal basis alongside CFKE=0 excitons for the hole added to a system of $28$ particles at $\nu = \frac{2}{5}$ as seen in Figure \ref{fig:hole-2-5-excitons-0-largest-size}. 

Having demonstrated the agreement between the spectral functions calculated from the minimal basis, extended minimal basis and ED, we can now use the minimal basis set to calculate the spectral function for larger systems not accessible with ED. 

\begin{figure*}
    \includegraphics[width=0.4\textwidth]{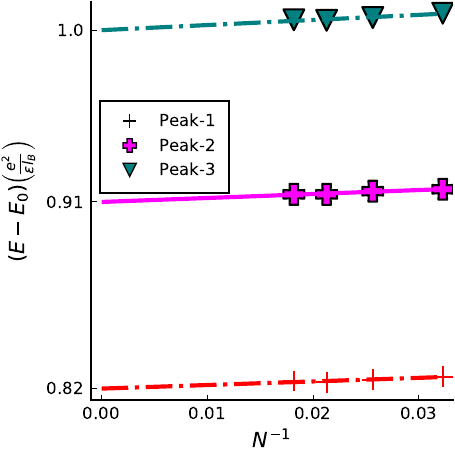}
    \includegraphics[width=0.4\textwidth]{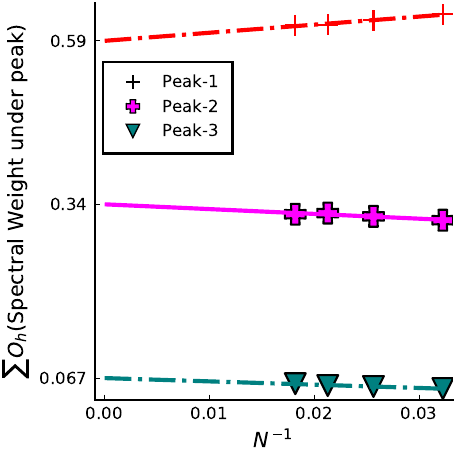}
    \caption{Thermodynamic extrapolation of the positions of the peaks (left) and the total spectral weight $\sum O_{h}$ underneath each peak (right) in the hole spectral function calculated using the minimal basis at $\nu = \frac{2}{5}$. Peak $x$ corresponds to the $x^{\rm th}$ lowest energy peak.}
    \label{fig:hole-2-5-thermodynamic-extrapolation}
\end{figure*}

Figure \ref{fig:hole-2-5-thermodynamic-extrapolation} shows the 
extrapolations to thermodynamic limit of the finite system estimates of peak positions (left) and the areas (right) i.e. the sum of spectral weights $\sum O_{h}$ under the spectral function peaks assuming a $1/N$ scaling of the leading finite size corrections.
\begin{figure*}
    \includegraphics[width=0.4\textwidth]{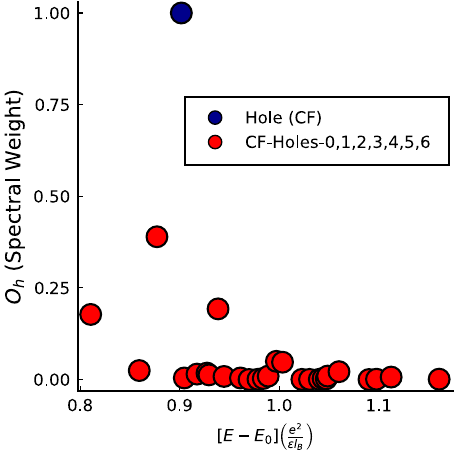}
    \includegraphics[width=0.4\textwidth]{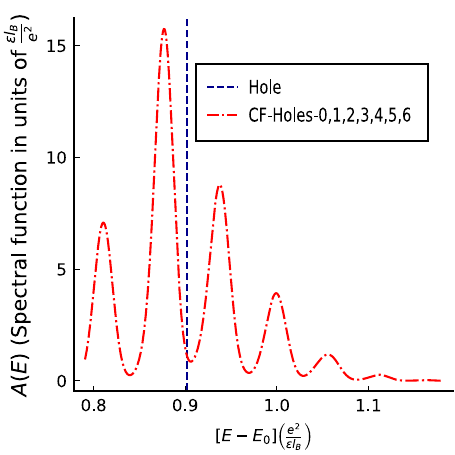}
    \caption{Spectral weights (left) and spectral function (right) calculated using the minimal basis for the hole added to an incompressible state of $54$ particles at $\nu = \frac{3}{7}$. The minimal basis at $\nu = \frac{3}{7}$ contains CF states with $7$ quasiholes distributed amongst the first $3$ CF LLs with CFKE ranging between $0$ and $6$. There are $39$ such states of which $27$ are at the same angular momentum as the hole.}
    \label{fig:hole-3-7-minimal-model}
\end{figure*}

In summary, we find one sharp peak in the hole spectral function at $\nu=1/3$, and two dominant peaks and one smaller peak at $\nu = \frac{2}{5}$. The number of peaks are in agreement with what is expected from the  minimal basis.
Number of minimal basis states increases with filling fraction. At $\nu = 3/7$ there are $39$ minimal basis states of which $27$ have the correct angular momentum quantum numbers and at $\nu = \frac{4}{9}$ there are $748$ minimal basis states of which $455$ have the correct angular momentum quantum numbers. As the number of minimal basis states increase, we do not expect the number of peaks to continue to match the number of basis states. Figure \ref{fig:hole-3-7-minimal-model} shows the spectral weights and spectral function calculated for the case of $\nu = \frac{3}{7}$ using the minimal basis, showing 4-6 peaks in the spectral function.

\section{Addition of an electron}\label{sec:electron-state}

This section provides a more complete account of the calculations for the electron spectral function leading to the thermodynamic results described in the main article.

The situation for the addition of an electron is more complicated. In this case, the wave function in the disk geometry is given by
\begin{multline}
c^{\dagger}(\vec{r}=0)|\Psi_{n/(2pn\pm 1}\rangle \propto \\
A \eta_{0,0}(\vec{r}_{N+1}) P_{\rm LLL}\Phi_n(\{ z_1,\cdots z_N\}) \prod_{j<k=1}^N(z_j-z_k)^2
\end{multline}
where $\eta_{0,0}(\vec{r})$ is the LLL wave function of an electron localized at the origin, and $A$ denotes antisymmetrization. This wave function does not have a natural representation in the CF form unlike the case of hole added state (Section \ref{supp_sec:holeDecomposition}), as the added electron is not Jastrow correlated with the rest of the electrons. We therefore expect a smaller spectral weight at low energies for an electron added into an FQH state(in comparison to the hole).

\begin{figure*}
    \includegraphics[width=0.4\textwidth]{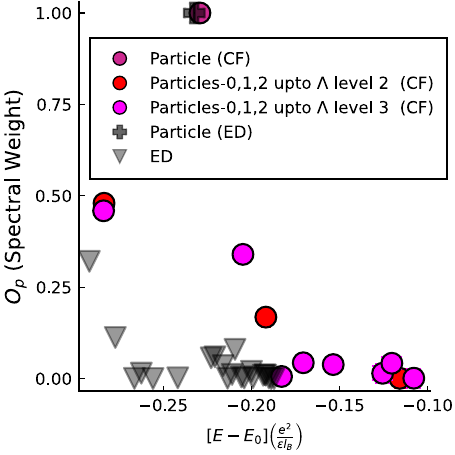}
    \includegraphics[width=0.4\textwidth]{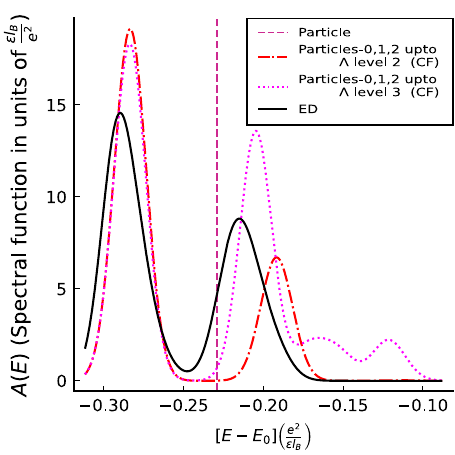}
    \caption{Spectral weights (left) and spectral function (right) calculated using different CF basis sets for the particle added to an incompressible state of $12$ particles at $\nu = \frac{1}{3}$. We have considered $2$ different CF basis sets in which the $3$ CF quasiparticles are distributed upto the $n=2$ and $n=3$ $\Lambda$ levels. The CF basis set with quasiparticles distributed upto the $n=3$ $\Lambda$ level provides the best approximation to the peaks from ED as seen in the right panel. Inclusion of states with higher CFKE or quasiparticles distributed in higher $\Lambda$ levels does not significantly increase the total spectral weight captured in the thermodynamic limit suggesting a marked asymmetry with the addition of a hole at $\nu = \frac{1}{3}$.}
    \label{fig:particle-1-3-ed-comparison}
\end{figure*}

\begin{figure*}
    \includegraphics[width=0.36\textwidth]{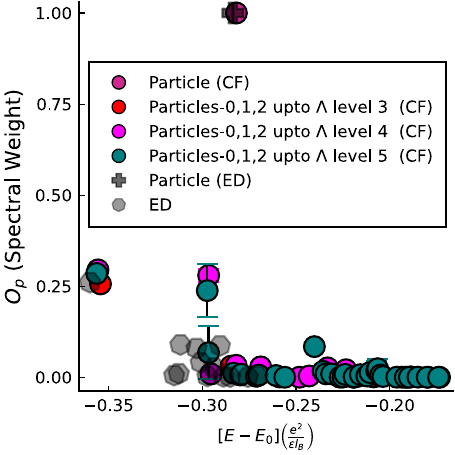}
    \includegraphics[width=0.36\textwidth]{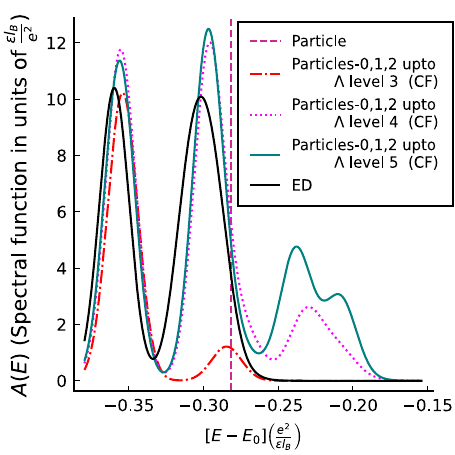}
    \caption{Spectral weights (left) and spectral function (right) calculated using different CF basis sets for the particle added to an incompressible state of $14$ particles at $\nu = \frac{2}{5}$. We have considered $3$ different CF basis sets in which the $5$ CF quasiparticles are distributed between the $n=3$, $n=4$ and $n=5$ $\Lambda$ levels. The CF basis set with quasiparticles distributed upto the $n=5$ $\Lambda$ level provides the best approximation to the peaks from ED as seen in the (right) spectral function plot. }
    \label{fig:particle-2-5-ed-comparison}
\end{figure*}

We can construct states with the right quantum numbers $L=L_z=Q$ with $2pn+1$ quasiparticles near the north pole to generate a basis to represent the electron added to the FQH state at filling fraction $\frac{n}{2pn+1}$. Figure 4 of the main text shows some of these states. There are an infinite number of such states if we allow the quasiparticles to occupy arbitrary $\Lambda$ levels and have arbitrary CFKE. A  upper cut off can be placed on the CFKE to get a finite basis set which may allow a calculation of the low energy spectral weights in the thermodynamic limit.

\begin{figure*}
    \includegraphics[width=0.4\textwidth]{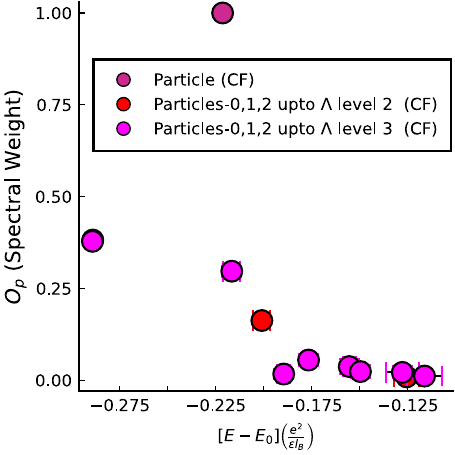}
    \includegraphics[width=0.4\textwidth]{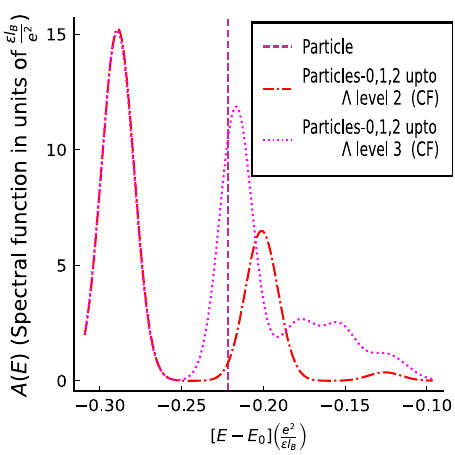}
    \caption{Spectral weights (left) and spectral function (right) calculated using different CF basis sets containing the $3$ quasiparticles distributed between the $n=2$ and $n=3$ $\Lambda$ levels to describe an electron added to the incompressible state of $32$ particles at filling $\nu = \frac{1}{3}$.}
    \label{fig:electron-added-to-13rd-33-particles}
\end{figure*}

\begin{figure*}
    \includegraphics[width=0.4\textwidth]{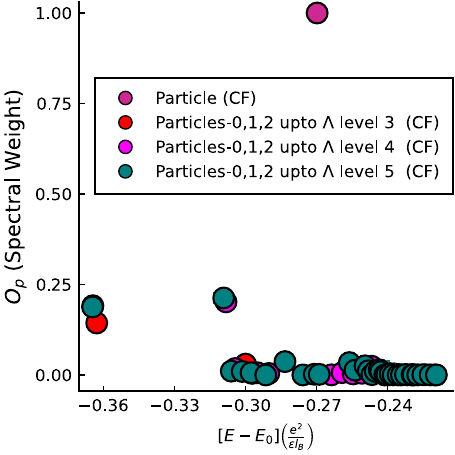}
    \includegraphics[width=0.4\textwidth]{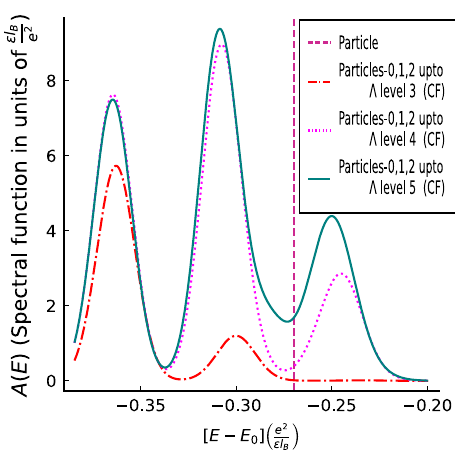}
    \caption{Spectral weights (left) and spectral function (right) calculated using different CF basis sets containing the $5$ quasiparticles distributed between the $n=3,4$ and $n=5$ $\Lambda$ levels to describe an electron added to the incompressible state of $28$ particles at filling $\nu = \frac{2}{5}$.}
    \label{fig:electron-added-to-25th-28-particles}
\end{figure*}

\begin{figure*}
    \includegraphics[width=0.4\textwidth]{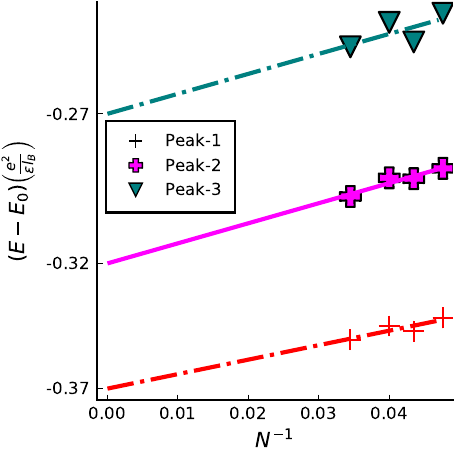}
    \includegraphics[width=0.4\textwidth]{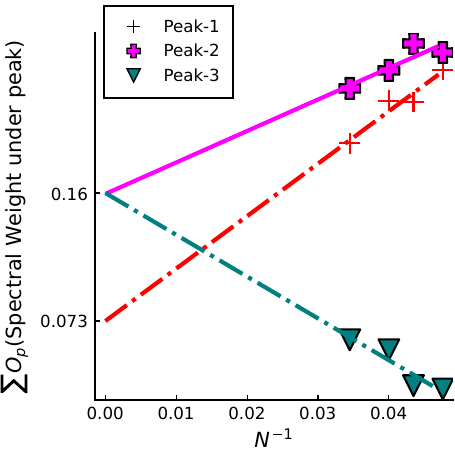}
    \caption{Thermodynamic extrapolation of the positions of the peaks (left) and the total spectral weight $\sum O_{p}$ underneath each peak (right) in the particle spectral function calculated using a CF basis containing the $5$ quasiparticles distributed upto the $n=5$ $\Lambda$ level at $\nu = \frac{2}{5}$.Peak $x$ corresponds to the $x^{\rm th}$ lowest energy peak.}
    \label{fig:particle-2-5-thermodynamic-extrapolation}
\end{figure*}

For the electron at $\nu = \frac{1}{3}$, we have considered several CF bases which differ in the upper cut-offs on the allowed $\Lambda$ levels, namely CF quasiparticles distributed up to the $n=2$ and $n=3$ $\Lambda$ levels with up to two units of CFKE (relative to the lowest energy basis state). The largest basis contains $16$ CF states.
For the electron at $\nu = \frac{2}{5}$, we have considered CF states with $5$ CF quasiparticles distributed upto $n=3,4$ and $n=5$ $\Lambda$ levels and up to two units of CFKE. The largest basis contains $60$ CF states. A comparison with exact diagonalization shows that these bases capture the behavior qualitatively and semi-quantitatively
(Fig. \ref{fig:particle-1-3-ed-comparison} and Fig. \ref{fig:particle-2-5-ed-comparison}). We have used the largest CF bases in each case i.e. the basis with upper cut-off of $n=3$ for the particle at $\nu = \frac{1}{3}$ and the basis with upper cut-off of $n=5$ for the particle at $\nu = \frac{2}{5}$ for our thermodynamic calculations(see for example Figure \ref{fig:electron-added-to-13rd-33-particles}, \ref{fig:electron-added-to-25th-28-particles} and \ref{fig:particle-2-5-thermodynamic-extrapolation}). These bases contain $\sim 76\%$ and $\sim 35\%$ of the spectral weight of the electron in the thermodynamic limits of $1/3$ and $2/5$ respectively.
Figure \ref{fig:particle-2-5-thermodynamic-extrapolation} shows the 
extrapolations to thermodynamic limit of the finite system estimates of peak positions (left) and the areas (right) i.e. the sum of spectral weights $\sum O_{p}$ under the spectral function peaks assuming a $1/N$ scaling of the leading finite size corrections.


\section{Decomposition of the hole in terms of CF wave functions\label{supp_sec:holeDecomposition}}
Here we show that the state formed by adding a hole at the north pole of the CF ground state at $n/(2pn+1)$ on the sphere can be written as a linear combination of simple CF states made of $2pn+1$ quasiholes and their simple neutral excitations near the north pole.

The FQHE ground state at electronic filling $\nu = \frac{n}{2np+1}$ with $N$ electrons which occur in $2Q=N/\nu - (n+2p)$ fluxes on the sphere can be written as
\begin{multline}
    \Psi_{\nu}(\Omega_{1},\dots,\Omega_{N}) = \LLL \phi_{n}(\Omega_{1},\dots,\Omega_{N}) \times \\\prod_{i<j}\left(u_{i}v_{j}-u_{j}v_{i}\right)^{2p}.
\end{multline}
Here $\phi_n$ is the Slater determinant representing the IQHE state of the $N$ composite fermions which fully occupy $n$ Landau levels in $2Q^{*} = 2Q-2p(N-1)$ fluxes. The Slater determinant $\phi_{n}$ is given by
\begin{widetext}
    \begin{equation}\label{eq:originalSphereSlaterDeterminant}
        \phi_{n}(\Omega_{1},\dots,\Omega_{N}) = \begin{vmatrix}
            Y_{Q^{\star},Q^{\star},Q^{\star}}(\Omega_{1}) & Y_{Q^{\star},Q^{\star},Q^{\star}}(\Omega_{2}) & \dots & Y_{Q^{\star},Q^{\star},Q^{\star}}(\Omega_{N}) \\
            Y_{Q^{\star},Q^{\star},Q^{\star}-1}(\Omega_{1}) & Y_{Q^{\star},Q^{\star},Q^{\star}-1}(\Omega_{2}) & \dots & Y_{Q^{\star},Q^{\star},Q^{\star}-1}(\Omega_{N}) \\
            \vdots & \vdots & \vdots & \vdots \\
            Y_{Q^{\star},Q^{\star}+1,Q^{\star}+1}(\Omega_{1}) & Y_{Q^{\star},Q^{\star}+1,Q^{\star}+1}(\Omega_{2}) & \dots & Y_{Q^{\star},Q^{\star}+1,Q^{\star}+1}(\Omega_{N}) \\
            \vdots & \vdots & \vdots & \vdots \\
            Y_{Q^{\star},Q^{\star}+n-1,-\left( Q^{\star}+n-1 \right)}(\Omega_{1}) & Y_{Q^{\star},Q^{\star}+n-1,-\left( Q^{\star}+n-1 \right)}(\Omega_{2}) & \dots & Y_{Q^{\star},Q^{\star}+n-1,-\left( Q^{\star}+n-1 \right)}(\Omega_{N})
        \end{vmatrix}
\end{equation}
\end{widetext}
It is sufficient to demonstrate the linear decomposition for the scenario where the hole is being added to the unprojected form of the CF ground state. This is because for a LLL electron annihilation operator $c$, we have $\LLL c \Psi^{\rm unproj}_{\nu} =  c \LLL\Psi^{\rm up}_{\nu}.$

Now we demonstrate that the state $c \Psi^{\rm unproj}_{\nu}$ can be expanded in a set of simple CF states. Consider the addition of a hole at the north pole $\omega\equiv (\theta=0,\phi)$. Following Eq.~\eqref{eq:electronRemovedState}, this is given by 
\begin{multline}
\Psi^{\rm unproj}_{\nu-e} = c\Psi^{\rm unproj}_{\nu}=\prod_{i<j=1}^{N-1}\left(u_{i}v_{j}-u_{j}v_{i}\right)^{2p}\times\\  \prod_{i=1}^{N-1}v_{i}^{2p} \times \phi_{n} (\Omega_{1},\dots\Omega_{N-1}, \omega)
\end{multline}
The quantity in the second line is antisymmetric in the electron coordinates $\{\Omega_,\dots \Omega_{N-1}\}$ and therefore can be expanded in Slater determinants of single particle angular momentum orbitals. This allows us to interpret the state $\Psi^{\rm up}_{\nu-e}$ as a linear combination of different CF states with fixed CF orbital occupancy. The CF orbitals that are occupied in these states depend on the Slater determinants in this expansion. We therefore study this Slater determinant expansion below.

We note that $\phi_{n} (\Omega_{1},\dots\Omega_{N-1}, \omega)$ can be written as a linear combination of Slater determinants of single particle angular momentum orbitals of $N-1$ electrons. 
\begin{widetext}
    \begin{align}
        \phi_{n}(\Omega_{1},\Omega_{2},\dots,\Omega_{N-1},\omega) & \propto \begin{vmatrix}
            Y_{Q^{\star},Q^{\star},Q^{\star}}(\Omega_{1}) & Y_{Q^{\star},Q^{\star},Q^{\star}}(\Omega_{2}) & \dots & 1 \\
            Y_{Q^{\star},Q^{\star},Q^{\star}-1}(\Omega_{1}) & Y_{Q^{\star},Q^{\star},Q^{\star}-1}(\Omega_{2}) & \dots & 0 \\
            \vdots & \vdots & \vdots & \vdots \\
            Y_{Q^{\star},Q^{\star}+1,Q^{\star}+1}(\Omega_{1}) & Y_{Q^{\star},Q^{\star}+1,Q^{\star}+1}(\Omega_{2}) & \dots & 0 \\
            Y_{Q^{\star},Q^{\star}+1,Q^{\star}}(\Omega_{1}) & Y_{Q^{\star},Q^{\star}+1,Q^{\star}}(\Omega_{2}) & \dots & 1 \\
            \vdots & \vdots & \vdots & \vdots \\
            Y_{Q^{\star},Q^{\star}+n-1,-\left( Q^{\star}+n-1 \right)}(\Omega_{1}) & Y_{Q^{\star},Q^{\star}+n-1,-\left( Q^{\star}+n-1 \right)}(\Omega_{2}) & \dots & 0
        \end{vmatrix} \label{eq:NorthPoleHoleSlaterDet}\\
        & = \sum_{i=1}^{n}c_{i}\tilde{\phi}_{i}(\Omega_{1},\dots,\Omega_{N-1})\label{eq:SingleHoleIQHExpansion}
    \end{align}
\end{widetext}
where we have used $Y_{Q,l,m}(\theta=0)\propto \delta_{m,Q}$ to simplify the last column of Eq.~\eqref{eq:originalSphereSlaterDeterminant} to the form in Eq. \eqref{eq:NorthPoleHoleSlaterDet}. Each term $\tilde{\phi}_{i}(\Omega_{1},\dots,\Omega_{N-1})$ in the expansion above is an IQH state with $n$ LLs fully filled but with a hole in the $m=Q^*$ orbital of the $i^{\rm th}$ LL. 
Now we consider the effect of multiplication of $\tilde{\phi}_i$ by $\prod_k v_k^{2p}$ which corresponds to multiplication of each occupied single particle wave function of $\tilde{\phi}_i$ by $v^{2p}\propto Y_{p,p,-p}$. Product of these monopole harmonics can be expanded in monopole harmonics with the combined monopole strength to get
\begin{multline}\label{eq:yqlmProductExpansion}
Y_{p,p,-p}(\Omega)Y_{Q^{\star},l,m}(\Omega) \\ = \sum^{l+p}_{L \geq L_{\rm min}(Q^*,p,l,m)}S(p,p,-p;Q^{\star},l,m;L)Y_{Q^{\star}+p,L,m-p} 
\end{multline}
where \[L_{\rm min}(Q^*,p,l,m)=\max[Q^{\star}+p,l-p,\absolutevalue{m-p}]. \]
The coefficients $S$ can be exactly written in terms of the Wigner 3j symbols. The constraints in the summation arise from the conditions for the Wigner 3j symbols to be non-zero. This expansion implies (1) a reduction of the azimuthal quantum number from $m$ to $m-p$, (2) an increase in the monopole strength from $Q^*$ to $Q^*+p$ and (3) mixing between LLs. For instance (for not too small systems):
\begin{align*}
    v^{2p}Y_{Q^{\star},Q^{\star},m} &= \alpha Y_{Q^{\star}+p,Q^{\star}+p,m-p} \\
    v^{2p}Y_{Q^{\star},Q^{\star}+1,m} &= \beta Y_{Q^{\star}+p,Q^{\star}+p,m-p}+ \gamma Y_{Q^{\star}+p,Q^{\star}+p+1,m-p} 
\end{align*}
for some constants $\alpha, \beta, \gamma$. We note the following points:
\begin{enumerate}
    \item The total flux seen by the CFs is increased to $Q^*+p$. This adds $2p$ quasiholes into every one of the $n$ LLs of the Slater determinants $\tilde{\phi}_i$. Together with the quasihole already in the the $i^{\rm th}$ LL of $\tilde{\phi}_i$, these states have $2pn+1$ quasiholes.
    \item Now we can ask what the momenta of these quasiholes are. Inspecting the expression in Eq. \eqref{eq:yqlmProductExpansion} we see that the azimuthal angular momenta of each orbital is decreased by $p$ units and the maximum posible azimuthal angular momentum of each LL is increased by $p$ units. This produces $2p$ quasiholes in each LL. This effect is further modified by the next observation.
    \item The LL index of the states on the right hand side varies between some $n_{\rm min}=L_{\rm min}-Q^*-p\geq 0$ and $l-Q^*$. This suggests that the orbitals from the LL $k$ are scattered  across orbitals of the same azimuthal angular momentum in the lower LLs (LL index less than or equal $k$). Since highest occupied orbital in $\tilde{\phi}_i$ is $n$, none of the CF orbitals above LL $n$ are occupied in the CF basis states.
\end{enumerate}
We now illustrate how a CF basis for the expansion of the hole added state can be constructed from considering the above rules. In particular we will specify the CF occupancy in these CF states.

For the case of $1/(2p+1)$, in which only the lowest LL is filled i.e. $n=1$, there is only one term in the expansion Eq. \eqref{eq:SingleHoleIQHExpansion}. The state $\tilde{\phi}_{i=0}$ is given by a CF configuration with one hole in the LLL at the $m=Q^*$ orbital. From the observation (2) above, multiplication by $v^{2p}$ expands the available orbitals in the CF configuration by addition of $2p$ orbitals as $Q^*$ is increased to $Q^*+p$. This results in a total of $2p+1$ quasiholes in the orbitals $Q^*+p,\dots Q^*-p$ of the lowest CF LL. Since only one LL is occupied in the incompressible state, there is no possibility of ``LL mixing" mentioned in point (3) above. This precisely gives the three quasihole state described in Fig.~1 (left panel) of the main text.

Now we consider the case of $2/5$ for which $n=2,p=1$. $\tilde{\phi}_{1}$ and $\tilde{\phi}_{2}$ contain one qh in the LLs $1$ and $2$ respectively. Addition of $2p=2$ QHs in each LL (point (2) above) results in the right panels of the two rows of Fig. 2 of the main text. Mixing of the LLs upto $n=2$ according to the point (3) above then produces the remaining two states of the minimal basis.

We make a conservative choice of $L_{\rm min}=Q^*+p$ (i.e. LLL) in our calculations. We note three points regarding this basis:
\begin{enumerate}
    \item The entire spectral weight of the hole can be captured by the specified basis.
    \item These basis states are themselves not energy eigenstates. Diagonalization within this minimal basis can produce approximate energy eigenstates which will necessarily contain the entire spectral weight of the hole. 
    \item A more accurate set of energy eigenstates containing the hole can be obtained by extending the variational space with excitons.
    \item Not all of these basis states will have the same $L^2$ quantum number as the hole added state ($L=Q$). Such linear combinations of these basis states with the wrong quantum numbers will be orthogonal to the hole added state and can be eliminated from the CF diagonalization calculation. 
\end{enumerate}

\end{document}